\documentclass[pre,showpacs,twocolumn]{revtex4}
\usepackage{graphicx} 
\usepackage{dcolumn}  
\usepackage{bm}       
\usepackage{amsmath}
\usepackage{amsfonts}
\usepackage{amssymb}

\begin{document}

\title{Quasicrystalline and rational approximant wave patterns in
hydrodynamic and quantum nested wells.}

\author{A. Baz\'an and M. Torres}
\affiliation{Instituto de F\'{\i}sica Aplicada, Consejo Superior de
  Investigaciones Cient\'{\i}ficas, Serrano 144, 28006 Madrid, Spain.}

\author{G. Chiappe, E. Louis and J.A. Miralles}
\affiliation{Departamento de F\'{\i}sica Aplicada and Unidad
  Asociada del Consejo Superior de Investigaciones Cient\'{\i}ficas,
  Universidad de Alicante, San Vicente del Raspeig, Alicante 03690,
  Spain.}

\author{J. A. Verg\'es}
\affiliation{Departamento de Teor\'{\i}a de la Materia Condensada,
  Instituto de Ciencia de Materiales de Madrid, Consejo Superior de
  Investigaciones Cient\'{\i}ficas, Cantoblanco, Madrid 28049, Spain.}

\author{Gerardo G. Naumis}
 \affiliation{Instituto de F\'{\i}sica, Universidad Nacional Aut\'onoma de
  M\'exico, Apartado Postal 01000, 76230 D.F., M\'exico .}

\author{J.L. Arag\'on}
 \affiliation{Centro de F\'{\i}sica Aplicada y Tecnolog\'{\i}a
  Avanzada, Universidad Nacional Aut\'onoma de
  M\'exico, Apartado Postal 1-1010, Quer\'etaro 76000, M\'exico.}

\begin{abstract}
The eigenfunctions of nested wells with incommensurate boundary geometry, in both hydrodynamic shallow water regime and quantum cases, are systematically and exhaustively studied in this letter. The boundary arrangement of the nested wells consist of polygonal ones, square or hexagonal, with a concentric immersed similar but rotated well or plateau. A rich taxonomy of wave patterns, such as quasicrystalline states, their crystalline rational approximants and some other exotic but well known tilings, is found in these mimicked experiments. To our best knowledge, these hydrodynamic rational approximants are presented here for the first time in a hydrodynamic-quantum framework. The corresponding statistical nature of the energy level spacing distribution reflects this taxonomy by changing the spectral types.
\end{abstract}


\maketitle

One would believe that a quasiperiodic wave pattern, that has an orientational order without periodic translational symmetry \cite{Janot}, must be associated to an external single connected boundary. In such a case, classical analogues which model features of quantum systems, and prove nontrivial properties of this last systems, have stirred interest. For example, acoustic \cite{Maynard1} and hydrodynamic quasicrystals \cite{Beloshapkin} were previously reported, and the existence of Bloch-like states has recently been proved in such systems \cite{Torres3}. In all of these results, there was an imposed global quasiperiodicity either by the boundary conditions or a dynamical source, which makes the appearance of such patterns not so unexpected. However, it is also possible to confine quasicrystalline hydrodynamic modes within an inner isolated region of a bigger surface, as it was made in Ref. \cite{Torres4}. In that work, the experiment was realized under a linear regime and thus the fluid "sees" a double concentric non-connected boundary. Hence, it was essentially different to other reported non-linear quasicrystalline Faraday wave patterns \cite{Alstrom} where the pattern is not conditioned by the shape of the boundary due to the non-linearity of the Faraday phenomenon. 

Although Ref. \cite{Torres4} provided new insights into the experimental generation of quasiperiodic patterns in the linear regimen, unfortunately it does not fulfills well the shallow water approximation \cite{Torres2}, \emph{i.e.}, in which the surface wavelength is much larger than the liquid depth, making impossible to state a suitable hydrodynamic-quantum analogy. Since the quantum analogs of these confined hydrodynamic modes may be of relevance to design quantum confinements echibiting quasiperiodic electronic states, or their rational approximants, here we systematically study the eigenfunctions corresponding to nested wells under the shallow water regime, and their corresponding quantum analogues. For this goal, we numerically solve the hydrodynamic equation, which in this case turns out to be the Helmholtz equation, and a thight binding approach is applied for the quantum case.

The nested wells of our study consist of a polygonal external boundary, square or hexagonal, with a concentric similar but incommensurately rotated well or plateau. In the case of the squares, the rotation angle between both domains is 45$^\circ$, and in the case of the hexagons, this angle is 30$^\circ$. The bottom of the vessel was covered with a shallow liquid layer of depth $h_1$, and the inner well or plateau with a depth $h_2$.

In the hydrodynamic study we use the equation: $\partial_t ^2 \eta (\mathbf{r},t) = \nabla \cdot \left[g h(\mathbf{r}) \nabla \eta (\mathbf{r},t) \right]$, \cite{Torres2} where $\eta$ is the wave amplitude, $g$ is the acceleration due to gravity and $h(\mathbf{r})$ is the depth field, being the corresponding Helmholtz standing wave equations $(\nabla ^2 + \omega^2/c_i^2) \Psi (\mathbf{r}) =0$;  $i=1,2$; where $\Psi (\mathbf{r})$ is the amplitude of the corresponding standing wave, with the above mentioned adequate regions, 1 (outer) and 2 (inner), and $c_i$ are the phase velocities, $\omega/k_i = (g h_i)^{1/2}$, $\omega$ is the angular frequency and $k$ is the wave number. 

We calculate the square nested well for a depth ratio $h_1 / h_2 = 0.9, 1.1, 0.77$, and $1.3$ respectively, with about 1700 eigenvalues and their corresponding eigenfunctions exhaustively checked for each case. The hexagonal nested well case is calculated for a depth ratio $h_1 / h_2 = 0.9$ and $1.1$, respectively, including about 1600 eigenvalues and their corresponding eigenfunctions in both spectra. Due to the high symmetry of the problem, a lot of doubly degenerate states appear. Any eigenfunction of spectra is available under request.

\begin{figure}[t!]
\begin{center}
\includegraphics[width=7.0cm]{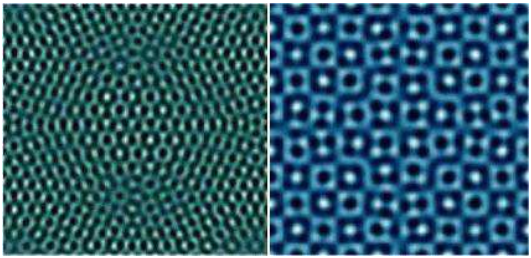}
\caption{Quantum octagonal quasicrystalline pattern (left) and its
$3/2$ rational approximant (right) in a double square nested well.
Simulation parameters are as follows. For the quasicrystalline
pattern, the well depth ratio is $1.16$ and the level energy, respect
to the $50$Hz ground state, is $-0.349174$. For the rational
approximant the well depth ratio is $1.1$ and the level energy is
$-0.2246$.}
 \label{fig:fig1}
\end{center}
\end{figure}

Since in the present r\'egime it is possible to establish an analogy with a quantum system, we also solve a tight-binding equation as described elsewhere \cite{Torres4,Cuevas}, with the same nested well geometry but providing it with a square atomic lattice. The correspondences with the hydrodynamic equations are the following:
\begin{eqnarray*}
E_{m,n} &=& 4 g h_{m,n} / a^2, \\
t_{m,n,m-1,n} &=& t_{m,n,m+1,n} = t_{m,n,m,n-1} \\
              &=& t_{m,n,m,n+1} = g h_{m,n} /a^2,
\end{eqnarray*}
where $E_{m,n}$ is the potential energy at the atomic site with integer lattice coordinates $(m,n)$, $h_{m,n}$ is the liquid depth at site $(m,n)$, $a^2 = L^2 / N^2$, where $L$ is the length of the external polygon side and $N$ is the discretization size (typically 201); and $t_{m, n, m^\prime, n^\prime}$ is the hopping integral between the nearest neighbour sites $(m,n)$ and $(m^\prime,n^\prime)$. Finally, in the present analogy, the dispersion relation corresponds to the equation: $E = \omega ^2 a^2/gh$. Spectra with 7.000 and 10.000 eigenvalues were respectively explored.

As we shall see, we present here quasicrystalline wave patterns and their lower rational approximants linked to an hydrodynamic-quantum analogy problem. In Fig. \ref{fig:fig1}, we show a quantum octagonal quasicrystalline pattern and its $3/2$ rational approximant one. The first pattern is obtained under quantum boundary conditions, \emph{i.e.}, Dirichlet conditions, and the second one, although it is also a quantum pattern, however it is obtained under hydrodynamic boundary conditions, \emph{i.e.}, Neumann conditions. Fig. \ref{fig:fig1} (left) conspicuously resembles to that pioneer experimental octagonal pattern early published in the very different non linear Faraday wave context by B. Christiansen, P. Alstrom and M. T. Levinsen \cite{Alstrom} (see Fig. 3(bottom) of this reference). A similar quasicrystalline octagonal pattern as that shown in Fig.\ref{fig:fig1} (left) is obtained under our present hydrodynamic framework for a depth ratio of $0.9$ and at an eigenfrequency of $f = \omega / 2 \pi = 3.02632$Hz.

\begin{figure}[t!]
\begin{center}
\includegraphics[width=7.0cm]{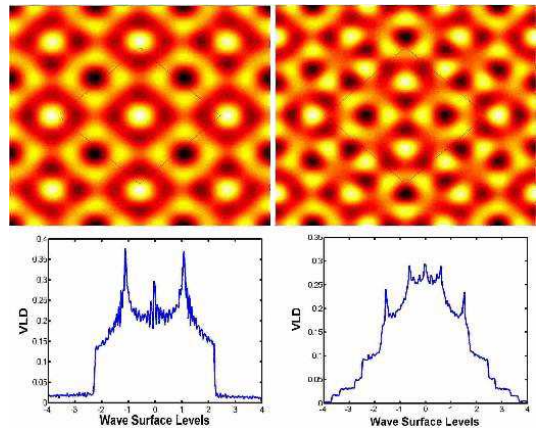}
\caption{Hydrodynamic rational approximants of the octagonal
pattern: $3/2$ (left top) and $7/5$ (right top). The edge-lengths of
the double square nested well are $80$ and $35$cm. In both cases the
depth ratio is $0.9$ and the frequencies for each case are $0.8376703$Hz
($3/2$ approximant) and $0.9505245$Hz ($7/5$ approximant). Below
each pattern, their corresponding vibrational level distribution (VLD) are shown.}
 \label{fig:fig2}
\end{center}
\end{figure}

The quasicrystalline octagonal pattern is generated by using linear combinations of two square lattice vector bases, shifted between them by an angle of $2 \tan ^{-1} \left( 2^{1/2} - 1 \right)$. By changing $2^{1/2}$ for its respective rational approximant numbers obtained from the continuous fraction expansion $2^{1/2} = [1; \bar{2}]$, the approximant patterns are generated. These wave vectors were accurately obtained for each eigenfunction by means of its corresponding pattern Fourier analysis. 

In Fig.\ref{fig:fig2} we also show both the hydrodynamic $3/2$ rational approximant pattern and the $7/5$ one with their corresponding vibrational level distributions $D(\eta_0)$ (VLD). The VLD is computationally obtained starting from the expression: $D(\eta_0) = \frac{1}{\Omega} \int_{\eta=\eta_0} \left( dl / | \nabla \eta (\mathbf{r}) | \right)$, where $\Omega$ is the area of the rational approximant unit cell. By replacing the $\mathbf{k}$-space by the real space, the above mentioned vibrational level distribution plays the same role in the corresponding eigenfunction as the well known density of states (DOS) in the whole spectrum \cite{Economou,Zaslavsky}.

\begin{figure}[t!]
\begin{center}
\includegraphics[width=7.0cm]{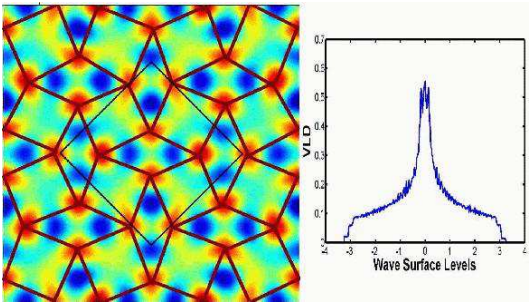}
\caption{Hydrodynamic pattern corresponding to the $\sigma$ phase
obtained with the same geometrical parameters of the Figure
\ref{fig:fig2} and with a depth ratio of $1.1$ at a frequency equal to
$0.7753178$Hz. Its corresponding VLD is shown at the right.}
 \label{fig:fig3}
\end{center}
\end{figure}

In Fig.\ref{fig:fig3}, we show an exotic pattern resembling the $\beta$-Mn structure (or $\sigma$-phase) which has been found coexisting as a crystalline approximant with the octagonal quasicrystalline phase \cite{Wang}. As Fourier analysed in this work, this pattern grows spontaneously in this mimicked experiment with the same wave vectors of the $3/2$ rational approximant one but introducing the phase $2 \arctan (1/2)$ in a couple of non-orthogonal vectors of the corresponding 4D basis. The phasing out transforms the VLD of the pattern, as shown in Fig. \ref{fig:fig3}(right), making it similar to that of the well known square lattice case \cite{Economou,Zaslavsky}. It is remarkable that the present hydrodynamic experiment can give insight about a basic crystallographic structural problem. To the authors best knowledge, this is the first time that the $\sigma$-phase is described in the density wave framework.

\begin{figure}[b!]
\begin{center}
\includegraphics[width=8.0cm]{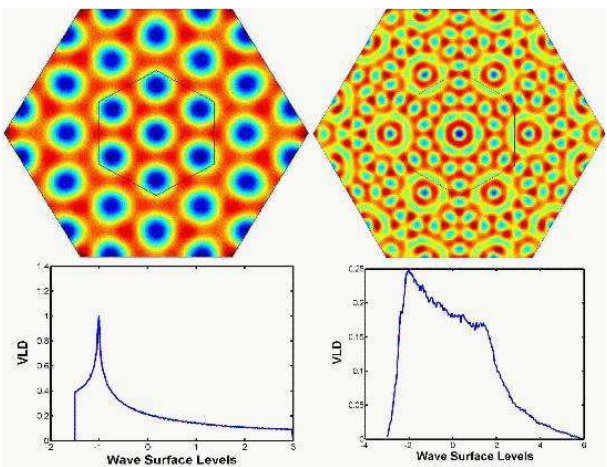}
\caption{Hydrodynamic triangular pattern (left) and the dodecagonal
$19/11$ rational approximant (right). The edge-lengths of the double
hexagonal nested well are $40$ and $17.5$cm respectively. The depth ratio and
frequency for each case are: $1.1$, $0.8155099$Hz (triangular) and
$0.9$, $2.2416974$Hz ($19/11$ approximant). Below each pattern,
their corresponding VLD are shown.}
 \label{fig:fig4}
\end{center}
\end{figure}

In a similar scenario, Fig.\ref{fig:fig4} shows two patterns generated in a hexagonal nested well and their corresponding level distributions. The quasicrystalline dodecagonal wave pattern can be also generated using two ternary wave vector sets shifted between them by an angle of $2 \tan ^{-1} \left( 2 - 3^{1/2} \right)$. Their rational approximant patterns appear when $3^{1/2}$ transforms to its corresponding rational approximant numbers. These rational numbers are obtained starting from the continuous fraction expansion $3^{1/2} = [1; \overline{1,2}]$.  Fig.\ref{fig:fig4} shows the $2$ and $19/11$ above mentioned rational approximant wave patterns. The VLD of the first pattern, or triangular pattern, is well known \cite{Economou,Zaslavsky}. The spectral problem and the rich taxonomy of eigenfunctions of this double-hexagonal experiment is similar to that previously described for the double square well.

\begin{figure}[t!]
\begin{center}
\includegraphics[width=7.0cm]{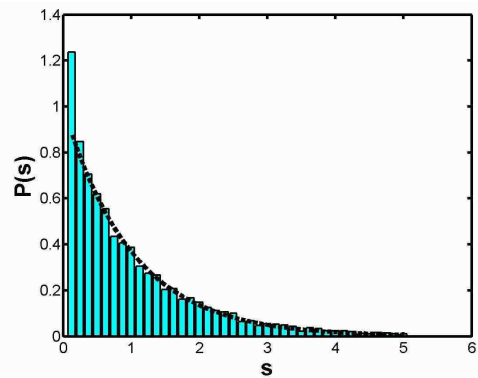}
\caption{Poisson distribution of nearest level spacing in the whole
level spectrum corresponding to the square symmetric nested wells case. Bars are experimental results and dashed line is the theoretical distribution.}
 \label{fig:fig5}
\end{center}
\end{figure}

\begin{figure}[t!]
\begin{center}
\includegraphics[width=6.0cm]{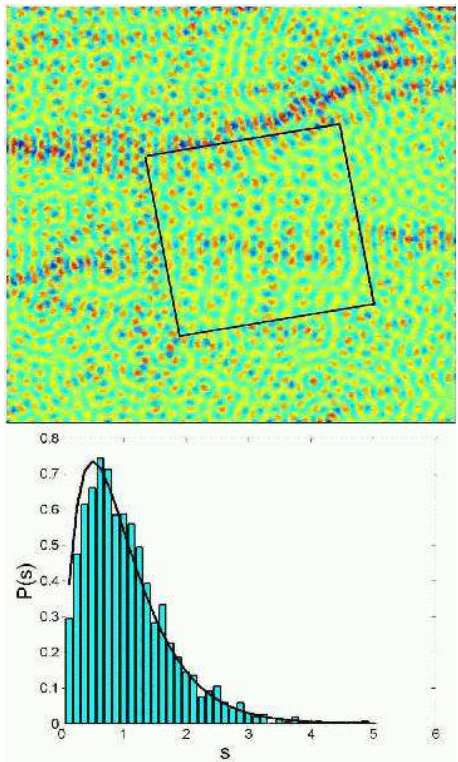}
\caption{Scarred chaotic wave pattern (top) obtained when the degeneracies 
are removed by rotating and shifting the inner well. An example is shown corresponding to a depth ratio of $0.77$ and at a eigenfrequency of $4.14917$Hz (bottom), and its Semi-Poisson distribution of 
energy spectral fluctuations. Continuous line represents the theoretical distribution 
and bars are experimental results.}
 \label{fig:fig6}
\end{center}
\end{figure}

As is well known \cite{Brody}, there is a relationship between the geometry of the boundary, the eigenfunction nature and the statistical properties of the spectrum. An integrable system, such as a polygonal billiard \cite{Wiersig}, has a statistical distribution of energy spectral fluctuations ($P(s)$), where $s$ is the distance between nearest energy levels, that fits the Poisson distribution. In Fig.\ref{fig:fig5}, we show that this is case for the $P(s)$ of the system under the symmetric configuration of the nested wells. If the symmetry of such system is broken, by slightly rotating and shifting the inner well, then all the inherent degeneracies are removed as it was made in a similar early study \cite{Richens}. In such a case, the eigenfunctions show "scarred" localized surface waves \cite{Kudrolli} as shown in Fig.\ref{fig:fig6}, and $P(s)$ becomes a semi-Poisson distribution.

We have shown that a geometrical arrangement of two nested wells, each of them with polygonal symmetry, can give a very rich and complex behavior even if a simple linear differential equation, valid for hydrodynamics and quantum mechanics, is used as physical description. Quasiperiodic, approximant and periodic wave patterns have been obtained in this work. The present approach gave a new insight on how to generate certain rational approximants in the density wave framework. The observed spectral fluctuations are consistent with the obtained wave patterns in both symmetrical and chaotic cases.

\begin{acknowledgments}
Technical support from S. Tehuacanero is gratefully acknowledged. This work has been partially supported by the Spanish MEC (FIS2004-03237 and MAT2002-04429) and the University of the Alicante, the Argentinean UBACYT (x210 and x447) and Fundacion Antorchas, and the Mexican DGAPA-UNAM (IN-108502-3) and CONACyT (D40615-F).
\end{acknowledgments}

\end{document}